\documentclass[final]{raa06}           
\usepackage{graphicx,times}             
\usepackage{natbib}
\usepackage{amssymb,amsmath}
\bibpunct{(}{)}{;}{a}{}{,}
\usepackage[colorlinks=true, citecolor=blue]{hyperref}%

\usepackage{graphicx,kantlipsum,setspace}
\usepackage{caption}
\usepackage{graphics,epsf}
\usepackage{amsmath}                
\usepackage{amsfonts}               
\usepackage{amssymb}                
\usepackage{epsfig}                 
\usepackage{appendix}
\usepackage{graphicx}
\usepackage{float}
\usepackage{color}
\usepackage{multirow}
\usepackage{colortbl}
\usepackage[para,online,flushleft]{threeparttable}
\usepackage{xcolor}

\hypersetup{citecolor=blue, 
            linkcolor=red, 
            menucolor=blue, 
            urlcolor=blue}  

\newcommand{\cm}{{~\rm cm}}

\newcommand{\km}{{~\rm km}}
\newcommand{\s}{{~\rm s}}

\begin{document}

   \title{The neutron star to black hole mass gap in the frame of the jittering jets explosion mechanism (JJEM)
}

   \volnopage{Vol.0 (20xx) No.0, 000--000}      
   \setcounter{page}{1}          

   \author{Noam Soker
    }

   \institute{Department of Physics, Technion, Haifa, 3200003, Israel;   {\it   soker@physics.technion.ac.il}\\
\vs\no
   {\small Received~~20xx month day; accepted~~20xx~~month day}}

\abstract{ I build a toy model in the frame of the jittering jets explosion mechanism (JJEM) of core collapse supernovae (CCSNe) that incorporates both the stochastically varying angular momentum component of the material that the newly born neutron star (NS) accretes and the constant angular momentum component and show that the JJEM can account for the $\simeq 2.5-5 M_\odot$ mass gap between NSs and black holes (BHs).  The random component of the angular momentum results from pre-collapse core convection fluctuations that are amplified by post-collapse instabilities. The fixed angular momentum component results from pre-collapse core rotation. 
For slowly rotating pre-collapse cores the stochastic angular momentum fluctuations form intermittent accretion disks (or belts) around the NS with varying angular momentum axes in all directions. The intermittent accretion disk/belt launches jets in all directions that expel the core material in all directions early on, hence leaving a NS remnant. Rapidly rotating pre-collapse cores form an accretion disk with angular momentum axis that is about the same as the pre-collapse core rotation. The NS launches jets along this axis and hence the jets avoid the equatorial plane region. In-flowing core material continues to feed the central object from the equatorial plane increasing the NS mass to form a BH. The narrow transition from slow to rapid pre-collapse core rotation, i.e., from an efficient to inefficient jet feedback mechanism, accounts for the sparsely populated mass gap. 
}

 \authorrunning{N. Soker}            
\titlerunning{Jets from a main sequence companion in CEE}  
   
      \maketitle

\section{Introduction} 
\label{sec:intro}

Observations indicate that there is a sparsely populated gap between the typical mass of neutron stars (NSs) and the masses of stellar-mass black holes (BHs), $M_{\rm BH} \ga 5 M_\odot$ (e.g.,  \citealt{Bailynetal1998, Ozeletal2010, Farretal2011, Kreidbergetal2012, LIGO3B}). This $\simeq 2.5-5 M_\odot$ gap is not completely empty (e.g., \citealt{Mrozetal2021, Lametal2022, LiGO3, LIGO3B}), and it is not clear yet how sparse is the population in the mass gap. 

If real, this gap in the mass distribution of core collapse supernova (CCSN) remnants is most likely related to the explosion mechanism of CCSNe. Recent theoretical studies consider two explosion mechanisms that utilise the gravitational energy that the collapsing core releases to explode the star. These are the delayed neutrino explosion mechanism (\citealt{BetheWilson1985}, with hundreds of studies since then, e.g., \citealt{Hegeretal2003, Janka2012, Nordhausetal2012, Mulleretal2019Jittering, BurrowsVartanyan2021, Fujibayashietal2021, Bocciolietal2022, Nakamuraetal2022}), and the jittering jets explosion mechanism (JJEM; \citealt{Soker2010, PapishSoker2011, GilkisSoker2015, Quataertetal2019, Soker2020RAA, ShishkinSoker2021, AntoniQuataert2022, Soker2022SNR0540, AntoniQuataert2023}). There are studies to account for the mass gap in the frame of the delayed neutrino explosion mechanism, e.g., \cite {Fryeretal2022} and \cite{Olejaketal2022}. 
In the present study I suggest an explanation in the frame of the JJEM that I base on the efficiency of the jet feedback mechanism. 

In the JJEM the newly born NS (or BH) launches jets as it accretes mass with stochastically varying specific angular momentum from the core (e.g., \citealt{Soker2010, PapishSoker2014Planar, GilkisSoker2015, Soker2019SASI, ShishkinSoker2022, Soker2022SNR0540, Soker2022Boosting}) or from the envelope (e.g., \citealt{Quataertetal2019, AntoniQuataert2022, AntoniQuataert2023}). The source of these angular momentum variations that lead to the formation of intermittent accretion disks or belts is the pre-collapse stochastic convection motion in the core or envelope that is amplified by instabilities between the newly born NS and the stalled shock at $\simeq 100 \km$ from the NS.
The intermittent accretion disks or belts launch jets with stochastically varying directions, i.e., jittering jets. 

Even when the specific angular momentum is somewhat below the limit to form an accretion disk, i.e., the accretion is through an accretion belt, the NS might launch jets \citep{SchreierSoker2016}. 
A support of this view comes from three-dimensional magneto-hydrodynamical simulations by \cite{Kaazetal2022} of a BH moving through a uniform magnetized medium. \cite{Kaazetal2022} find that even that the initial angular momentum of the accreted gas is zero the BH launches strong jets. The necessary condition in their simulations is that the magnetic fields are sufficiently strong. Since the jittering jets last for only few seconds and are generally not relativistic jets, the neutrino emission in the JJEM is as that in the neutrino-driven explosion mechanism and not as expected in the case of choked relativistic jets (e.g., \citealt{Guettaetal2023}). In case of BH formation, in the JJEM the jets might become relativistic and then have neutrino emission as in choked gamma-ray bursts (as calculated by, e.g., \citealt{Heetal2018, Fasanoetal2021}).   

There are some fundamental differences between the JJEM and many papers that study jet-driven explosions that operate only for rapidly rotating pre-collapse cores and therefore the jets that the newly born NS or BH launch have a fixed axis (e.g., \citealt{Khokhlovetal1999, Aloyetal2000, MacFadyenetal2001, Maedaetal2012, LopezCamaraetal2013, BrombergTchekhovskoy2016,  Nishimuraetal2017, WangWangDai2019RAA, Grimmettetal2021, Perleyetal2021, ObergaulingerReichert2023}). These differences are as follows (e.g., \citealt{Soker2022Rev}). (1) The JJEM operates in a jet negative feedback mechanism. By this the JJEM accounts for explosion energies that are several times the binding energy of the ejected mass. (2) The JJEM asserts that jets explode most, and possibly all, CCSNe and for all pre-explosion rotation rates of the cores. (3) According to the JJEM there are no failed CCSNe. All massive stars explode. 

\cite{ShishkinSoker2022} estimate the NS mass in cases of no pre-collapse core rotation, when the jet feedback mechanism during CCSN explosions is very efficient. The remnants are NSs. In the present study I include core rotation within a toy model (section \ref{sec:ToyModel}). I then use this toy model to account for the mass gap (sections \ref{sec:Feedback} and \ref{sec:Distribution}). I summarize this study in section \ref{sec:Summary}. 

\section{The assumptions of the toy model} 
\label{sec:ToyModel}

In the JJEM of CCSNe that have no pre-collapse core rotation the specific angular momentum of the gas that the newly born NS accretes varies stochastically. The seeds of these variations come from the convective motion in the pre-collapse core (e.g., \citealt{GilkisSoker2014, GilkisSoker2016, ShishkinSoker2021}) or envelope (e.g., \citealt{Quataertetal2019}). Post-shock instabilities behind the stalled shock inside a radius of $r \simeq 100 \km$ (\citealt{Soker2019SASI, Soker2019arXiv}) amplify these seeds to much larger values. Such an instability is likely to be the spiral standing accretion shock instability (spiral SASI; for simulations of the spiral SASI see, e.g.,  \citealt{Andresenetal2019, Walketal2020, Nagakuraetal2021, Shibagakietal2021}).  This phase of accretion that leads to the formation of intermittent accretion disks starts after the formation of the stalled shock. The baryonic mass inside the stalled shock at that time is $\approx 1-1.2 M_\odot$ (e.g., \citealt{Jankaetal2007}). 
  
There are $\simeq {\rm few}$ to $\simeq 30$ jet-launching episodes during an explosion with no pre-collapse core rotation, each lasts for a time period of $\simeq 0.01-0.1 \sec$, and the typical terminal velocity of the jets is $\simeq 10^5 \km \s^{-1}$ (\citealt{PapishSoker2014a}; neutrino observations limit the jets in most cases, excluding gamma ray bursts and similar transients, to be non-relativistic; \citealt{Guettaetal2020}). Each accretion disk of an episode has a mass of $\approx 10^{-2} M_\odot$ and the jets of each episode carry $\approx 10\%$ of this mass.  Although the asymmetrical accretion process onto the NS accompanied by rotation can lead to gravitational wave emission (e.g., \citealt{Dall'OssoStella2007, Menonetal2023}), based on the results of \cite{Gottliebetal2023} I expect that in the JJEM the jittering jets are the major source of gravitational waves.  

\begin{figure}[t]
	\centering
\includegraphics[trim=5.0cm 15.5cm 1.0cm 3.0cm ,clip, scale=0.99]{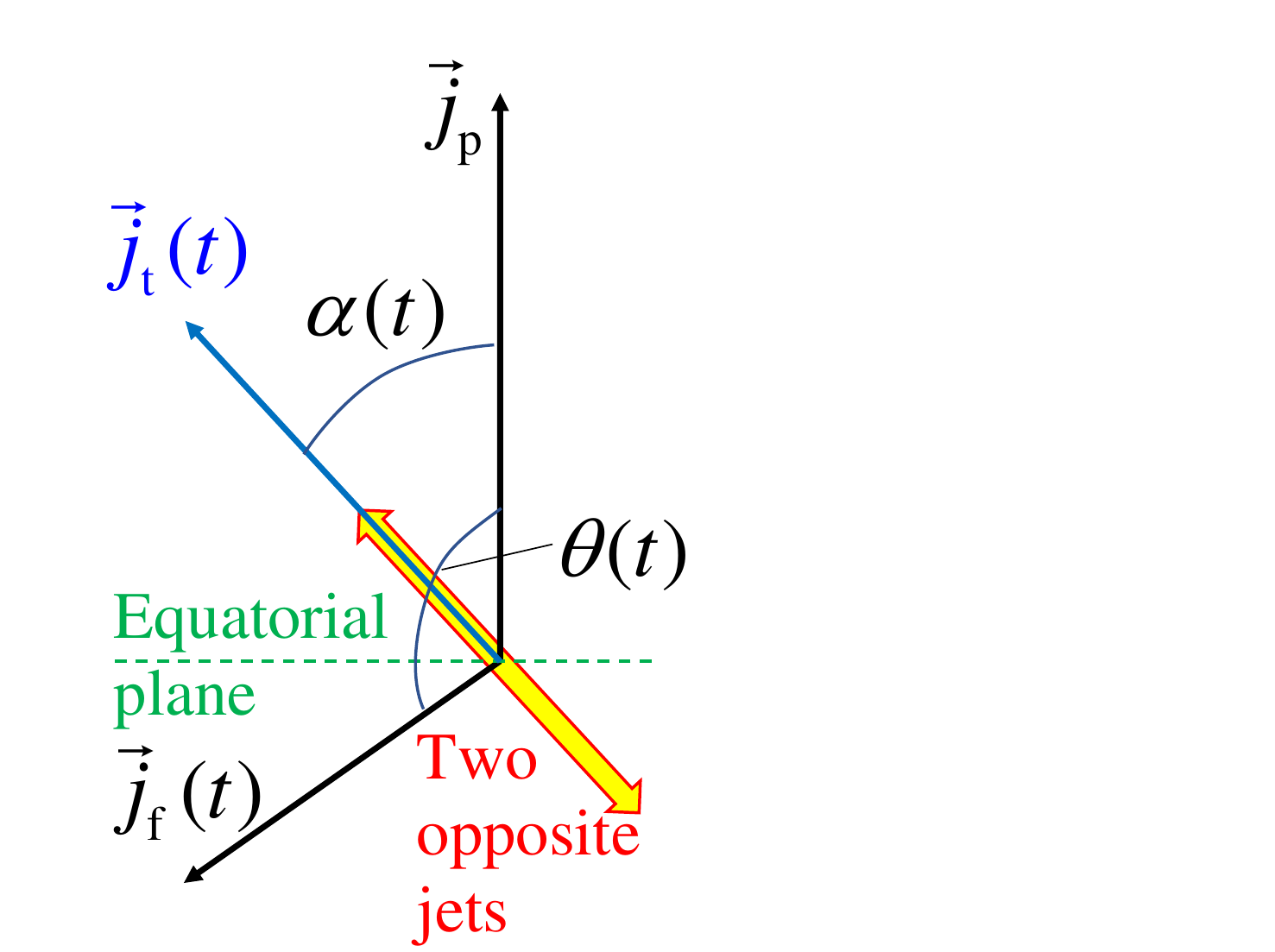} 
\caption{A schematic drawing of the sum of the stochastic component of the specific angular momentum $\overrightarrow{j_{\rm f}}(t)$, which results from pre-collapse core convection amplified by instabilities and changes rapidly and stochastically with time, and of the fixed component $\overrightarrow{j_{\rm p}}$ that results from the pre-collapse core rotation (see table \ref{TAB:Table1}). The NS launches two opposite jets for a short time along the momentarily direction of the axis of $\overrightarrow{j_{\rm t}}(t)=\overrightarrow{j_{\rm f}}(t)+\overrightarrow{j_{\rm p}}$. The equatorial plane is that of the pre-collapse core rotation. 
}
\label{Fig:Schamatic}
\end{figure}

In this study I build a very simple version of this complex explosion mechanism (a toy model). I assume that all specific angular momentum fluctuations of the gas that the newly born NS (a NS age of less than a few seconds) accretes after amplification by post-shock instabilities have the same magnitude of $j_{\rm f}$ and stochastically direction variations; `f' stands for fluctuating directions. I include the pre-collapse core rotation as an additional constant specific angular momentum $\overrightarrow{j_{\rm p}}$; `p' stands for pre-collapse rotation. 
I take the angle between $\overrightarrow{j_{\rm f}}(t)$ and the constant direction of $\overrightarrow{j_{\rm p}}$ to be $\theta(t)$, as I draw schematically in Fig. \ref{Fig:Schamatic}. Fig. \ref{Fig:Schamatic} is drawn in the momentarily plane of the two angular momenta $\overrightarrow{j_{\rm f}}(t)$ and $\overrightarrow{j_{\rm p}}$. Each new fluctuation of $\overrightarrow{j_{\rm f}}(t)$ defines a different plane.  

Under the assumption of fully stochastic variations of the directions of $\overrightarrow{j_{\rm f}}$ the probability for an angle $\theta (t)$ is 
\begin{equation}
f(\theta) d \theta = \sin \theta d \theta. 
\label{eq:ftheta}
\end{equation}
The intermittent accretion disk launches the jets along the new angular momentum direction $\overrightarrow{j_{\rm t}}=\overrightarrow{j_{\rm f}}+ \overrightarrow{j_{\rm p}}$.
The angle $\alpha(t)$ of $\overrightarrow{j_{\rm t}}$ with respect to $\overrightarrow{j_{\rm p}}$ is given by 
\begin{equation}
\tan \alpha = \frac{j_{\rm f} \sin \theta}{j_{\rm f} \cos \theta + j_{\rm p}} = 
\frac{\sin \theta}{\cos \theta + \beta_{\rm p}} ,  
\label{eq:alpha1}
\end{equation}
where 
\begin{equation}
\beta_{\rm p} \equiv \frac{j_{\rm p}}{j_{\rm f}}. 
\label{eq:betap}
\end{equation}

I search for the maximum angle of the jets' axis with respect to $\overrightarrow{j_{\rm p}}$. This would be the maximum value of $\sin \alpha$ as $0 \le \alpha \le 180^\circ$.
The condition $d \sin \alpha / d \theta =0$ yields a value $\theta_{\rm A}$ that gives the maximum angle $\alpha_{\rm A}$ for the angle of the specific angular momentum of the accreted mass with respect to the pre-collapse angular momentum
\begin{equation}
\cos \theta_{\rm A} =
\left\{
	\begin{array}{ll}
		-\beta_{\rm p}  & \mbox{if } ~ \beta_{\rm p} \le 1  \\
		-\frac{1}{\beta_{\rm p}}  & \mbox{if }~  \beta_{\rm p} > 1 .
	\end{array}
\right.
\label{eq:CosThetaMax}
\end{equation}
The corresponding maximum value of $\sin \alpha$ is 
\begin{equation}
\sin \alpha_{\rm A} =
\left\{
	\begin{array}{ll}
		1  & \mbox{if } ~ \beta_{\rm p} \le 1  \\
		\frac{1}{\beta_{\rm p}}  & \mbox{if } ~ \beta_{\rm p} > 1 .
	\end{array}
\right.
\label{eq:SinAlpha1}
\end{equation}

For the launching of jets I take the condition that the specific angular momentum of the accreted mass should be larger than a minimum limit of $j_{\rm L}$. Namely, 
\begin{equation}
\vert \overrightarrow{j_{\rm f}}+ \overrightarrow{j_{\rm p}} \vert \ge j_{\rm L} \simeq 2 \times 10^{16} \cm^2 \s^{-1}
\quad {\rm for~jets.}
\label{eq:Jlimit1}
\end{equation}
With the help of Fig. \ref{Fig:Schamatic} condition (\ref{eq:Jlimit1}) reads  
\begin{equation}
\sqrt{({j_{\rm f}} \cos \theta + {j_{\rm p}} )^2 + ({j_{\rm f}} \sin \theta)^2} \ge j_{\rm L} ,
\label{eq:Jlimit2}
\end{equation}
which yields 
\begin{equation}
\cos \theta \ge \cos \theta_{\rm mL} = \frac{1}{2 \beta_{\rm p}}\left[ \left( \frac{j_{\rm L}}{j_{\rm f}} \right)^2 - \beta^2_{\rm p} -1 \right]. 
\label{eq:Jlimit3}
\end{equation}
If this equation gives $\cos \theta_{\rm mL} < -1$ then $\cos \theta_{\rm mL} = -1$ and this condition does not limit the maximum angle $\alpha$.   
If the value of $\theta_{\rm A}$ as given by equation (\ref{eq:CosThetaMax}) does not fulfil condition (\ref{eq:Jlimit3}), then I take the maximum angle $\alpha$ to be the one given from equation (\ref{eq:alpha1}) for $\theta_{\rm mL}$. The final expression for the maximum angle of the jets relative to $\overrightarrow{j_{\rm p}}$ is
\begin{equation}
\alpha_{\rm max} = \min [\alpha_{\rm A}, \alpha(\theta_{\rm mL})]. 
\label{eq:Alphafinal}
\end{equation}

I summarize the different variables of the toy model in Table \ref{TAB:Table1}. 
\begin{table*}[]
\begin{tabular}{|p{3.8cm}|p{3.8cm}|p{3.8cm}|p{3.8cm}|}
\hline
Variable & Source &  Properties & Toy model\\ 
\hline
$\overrightarrow{j_{\rm f}} (t) $: Fluctuating specific angular momentum component.  &  Pre-collapse core convection amplified by post-shock instabilities. &  Stochastic magnitude and direction variations; vary  between jets-launching episodes. & Assumes constant magnitude $j_{\rm f}$ and stochastic direction (equation \ref{eq:ftheta}).\\ 
\hline
$\overrightarrow{j_{\rm p}}$: Specific angular momentum with a constant direction.  &  Pre-collapse core rotation. & Increases as accretion proceeds because ${j_{\rm p}} (r)$ increases with pre-collapse radius in the core. & Assumes constant $\overrightarrow{j_{\rm p}}$ in each CCSN. Different CCSNe have different ${j_{\rm p}}$ values. \\ 
\hline 
$\overrightarrow{j_{\rm t}}$: The specific angular momentum of the accreted gas. & $\overrightarrow{j_{\rm t}} = \overrightarrow{j_{\rm f}} + \overrightarrow{j_{\rm p}}$ & Varies between jet-launching episodes. An angle $\alpha$ to $\overrightarrow{j_{\rm p}}$ (Fig. \ref{Fig:Schamatic}). & The direction of $\overrightarrow{j_{\rm t}}$ is the axis of the two opposite jets, if launched. \\ 
\hline 
${j_{\rm L}}$: The minimum specific angular momentum of the accreted gas to launch jets.  & Physics of jet's launching. & ${j_{\rm L}} \simeq 2 \times 10^{16} \cm^2 \s^{-1}$ (equations \ref{eq:Jlimit1}-\ref{eq:Jlimit3}).  & According to the jittering jet explosion mechanism ${j_{\rm L}}< j_{\rm f}$. \\ 
\hline
$\alpha_{\rm max}$: Maximum possible angle of jets relative to $\overrightarrow{j_{\rm p}}$ (Fig. \ref{Fig:Schamatic}). &  The direction of $\overrightarrow{j_{\rm t}}$ (equation \ref{eq:SinAlpha1}) and the demand ${j_{\rm t}} \ge {j_{\rm L}}$ (equation \ref{eq:Jlimit3}). & For rapid pre-collapse core rotation $\alpha_{\rm max} < 90^\circ$ (equation \ref{eq:SinAlpha1}; Fig. \ref{Fig:MaxAngle}). & When $\alpha_{\rm max} < 90^\circ$ the jet feedback mechanism is not efficient and equatorial inflow turns the NS to a BH. \\
\hline
\end{tabular}
\caption{Summary of variables of the toy model (see also Fig. \ref{Fig:Schamatic}). 
}
\label{TAB:Table1}
\end{table*}
\normalsize
     
The minimum value to form an accretion disk around a NS of mass $M_{\rm NS}=1.4M_\odot$ and a radius of $R_{\rm NS}=12 \km$ is $2.1 \times 10^{16} \cm^2 \s^{-1}$. The radius of the very young NS is somewhat larger. However, an accretion belt, i.e., with somewhat sub-Keplerian specific angular momentum, can also launch jets. For that reason I use for $j_{\rm L}$ the value as in equation (\ref{eq:Jlimit1}). 
The pre-collapse radius of the accreted gas during the launching of jittering jets is $R_{\rm pc} \simeq 2000-5000 \km \s^{-1}$. The maximum allowed specific angular momentum of a test particle at these radii is $j_{\rm p,max} = 1.9\times 10^{17} (R_{\rm pc}/2000 \km)^{1/2} \cm^2 \s^{-1} \simeq 10 j_{\rm L}$. The relevant range of values for the present study is therefore $0 \le (j_{\rm p}/j_{\rm L})  \la 10$. Under the assumption of the jittering jets explosion mechanism the fluctuation alone can launch jets even when $j_{\rm p}=0$. I also demand therefore that $j_{\rm L} \le j_{\rm f}$.   

The toy model assumes that jets cannot expel mass perpendicular to their axis. 
\cite{Gottlieb_etal_2022_wobbly} conduct simulations of relativistic jets from an already formed BH of $4 M_\odot$ and find that strong fixed-axis jets might remove some mass even from the equatorial plane. On the other hand, simulations that are relevant to the jittering jets explosion mechanism (\citealt{PapishSoker2014a, PapishSoker2014Planar}) show that non-relativistic jets (as expected here) from a NS do not remove mass from the equatorial plane.

\section{The steep transition from efficient to inefficient jet feedback} 
\label{sec:Feedback}

I consider the launching of two opposite jets in each jet-launching episode along the total angular momentum axis, i.e., along $\overrightarrow{j_{\rm t}}(t)=\overrightarrow{j_{\rm f}}(t)+ \overrightarrow{j_{\rm p}}$. The total specific angular momentum $\overrightarrow{j_{\rm t}}(t)$ has an angle $\alpha(t)$ with respect to the pre-collapse angular momentum $\overrightarrow{j_{\rm p}}$ that is given by equation (\ref{eq:Alphafinal}). The angle $\phi(t)$ around the axis $\overrightarrow{j_{\rm p}}$ randomly changes between different jet-launching episodes in the range of $0 - 360^\circ$. Table
 \ref{TAB:Table1} summarises the variables of the toy model. 

I present the value of $\alpha_{\rm max}$ as a function of $j_{p}/j_{\rm L}$ and for three values of the fluctuating specific angular momentum component $j_{\rm f}$. When $\beta_{\rm p} = j_{\rm p}/j_{\rm f} > 1$ and condition (\ref{eq:Jlimit3}) is fulfilled, equation (\ref{eq:SinAlpha1}) shows that 
$\sin \alpha_{\rm max}=(j_{\rm f}/j_{\rm L})(j_{\rm p}/j_{\rm L})^{-1}$.
\begin{figure}[t]
	\centering
\includegraphics[trim=3.2cm 8.2cm 3.0cm 6.2cm ,clip, scale=0.60]{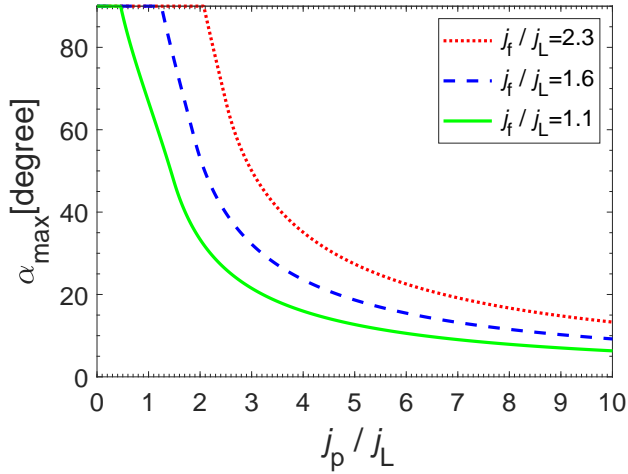} 
\caption{The maximum possible angle of the jets' axis with respect to the direction of the pre-collapse rotation axis of the core (the direction of $\overrightarrow{j_{\rm p}}$) as function of $j_{\rm p}/j_{\rm L}$, where $j_{\rm p}$ is the pre-collapse specific angular momentum in the core and $j_{\rm L}$ is the minimum specific angular momentum required to launch jets. The three lines are for different values of the specific angular momentum amplitude of the fluctuating component $j_{\rm f}$ that results from pre-collapse convection motion amplified by instabilities. Note that for $j_{\rm p}/j_{\rm f} \gtrsim 0.5$ (solid-green line) to $j_{\rm p}/j_{\rm f} \gtrsim 0.9$ (dotted-red line) no jets are launched in the equatorial plane ($\alpha = 90^\circ$).
}
\label{Fig:MaxAngle}
\end{figure}

The implication of $\alpha_{\rm max} < 90^\circ$ is that there are no jets in the equatorial plane. In that case the jets do not directly expel core material from the equatorial plane, allowing the equatorial accretion process to continue. This leads to an inefficient jet feedback mechanism where the accretion process and jet-launching last for a long time, up to minutes to even hours and more. An inefficient jet feedback mechanism might lead to very energetic explosion (e.g., \citealt{Gilkisetal2016Super}). This prolonged accretion turns the NS to a BH of several solar masses and more. 

Each of the three lines in Fig, \ref{Fig:MaxAngle} presents the same qualitative behavior of three regimes. The regime are determined by the values of  $j_{\rm p}/j_{\rm f}=(j_{\rm p}/j_{\rm L})/(j_{\rm f}/j_{\rm L})$, where the value of the nominator on the right hand side is from the  horizontal axis and the value of the denominator is according to the inset.  
\begin{enumerate}
    \item $j_{\rm p} \la (0.5-1)j_{\rm f}$. For slowly rotating pre-collapse core where the specific angular momentum in the core is significantly smaller than the fluctuating specific angular momentum component of the accreted gas the maximum angle is  $\alpha_{\rm max} = 90^\circ$. The jets are launched in all directions and  the jet feedback mechanism is efficient. The jets explode the star and terminate accretion in few seconds or less. As a result of that the remnant is a NS. 
    \item $j_{\rm p} \simeq j_{\rm f}$. Within this narrow range of $j_{\rm p}$ the value of $\alpha_{\rm max}$ drops from $90^\circ$ to much lower values, i.e., $\alpha_{\rm max} < 70^\circ$. I further discuss this regime below.  
    \item $j_{\rm p} > j_{\rm f}$. For rapidly rotating pre-collapse core there are no jets in the equatorial plane, i.e., $\alpha_{\rm max} \ll 90^\circ$. This limit occur for $j_{\rm p} > (\sin \alpha_{\rm max})^{-1} j_{\rm f}$, or $j_{\rm p} =1.15 j_{\rm f}$ for not too small $j_{\rm L}$. In this regime the jets are inefficient in expelling core material from the equatorial plane and the outcome is a long lasting accretion phase that turns the newly born NS into a BH with several solar masses or more.          
\end{enumerate} 

I consider cases in the middle regime of $j_{\rm p} \simeq j_{\rm f}$. The accretion processes via accretion disks or belts start when $j_{\rm p} \simeq j_{\rm f}$. The maximum angle might be close to $90^\circ$, i.e., $\alpha_{\rm max} \simeq 90^\circ$. In this case the accretion proceed to make a NS of mass $>1.4 M_\odot$. This in principle can form objects in the gap, i.e., in the range of $\simeq 2.5-5 M_\odot$. However, as the mass accretion continues and gas from further out in the core is accreted the specific angular momentum $j_{\rm p}$ increases because the specific angular momentum in the pre-collapse core increase. For example, for a solid-body rotation $j_{\rm p}  \propto r^2$. The increase in the value of $j_{\rm p}$ decreases the value of $\alpha_{\rm max}$ and makes the feedback process less efficient even. 

Overall, not only the range of pre-collapse core rotation that allows for remnant mass in the gap to form is relatively narrow ($j_{\rm p} \simeq j_{\rm f}$), but in  addition as accretion in this regime starts the value of $j_{\rm p}$ rapidly increases to higher values of $j_{\rm p}> j_{\rm f}$ that lead to BH formation. Overall, only a very small fraction of the remnants will be in the mass gap. 

\section{The distribution of the jets' angles} 
\label{sec:Distribution}
I turn to find the distribution function of the angle $\alpha$ between the jets and the fix axis along $\overrightarrow{j_{\rm p}}$. When $j_{\rm p}=0$ the jets are launched along the momentarily direction of $\overrightarrow{j_{\rm f}}$. In this case $\alpha=\theta$ (see Fig. \ref{Fig:Schamatic}) and the distribution is according to equation (\ref{eq:ftheta}) with $F_0(\alpha)=f(\theta)$. Namely, 
\begin{equation}
F_0(\alpha)~ d \alpha = \sin \alpha ~d \alpha \quad {\rm for} \quad j_{\rm p}=0. 
\label{eq:Falpha0}
\end{equation}
For the general case of $j_{\rm p}>0$  
\begin{equation}
F(j_{\rm p}, \alpha) = f(\theta) \left\vert \frac {d \theta} {d \alpha} \right\vert= 
\sin \theta
\frac{1+2 \beta_{\rm p} \cos \theta + \beta^2_{\rm p}} 
{1+\beta_{\rm p} \cos \theta }  
\label{eq:Falpha}
\end{equation}
where $d \theta /d \alpha$ is calculated from equation (\ref{eq:alpha1}). 
For $\beta_{\rm p}>1$ two angles $\theta_1$ and $\theta_2$ give the same angle $\alpha$. Namely, there are two contributions to $F(j_{\rm p}, \alpha)$. 

Another condition is that the specific angular momentum of the accreted gas obeys condition (\ref{eq:Jlimit1}), i.e., that the values of $\theta$ obeys condition (\ref{eq:Jlimit3}). If this condition is violated I set $F(j_{\rm p}, \alpha)=0$. 

\begin{figure}[t]
	\centering
\includegraphics[trim=3.2cm 8.2cm 3.0cm 7.2cm ,clip, scale=0.60]{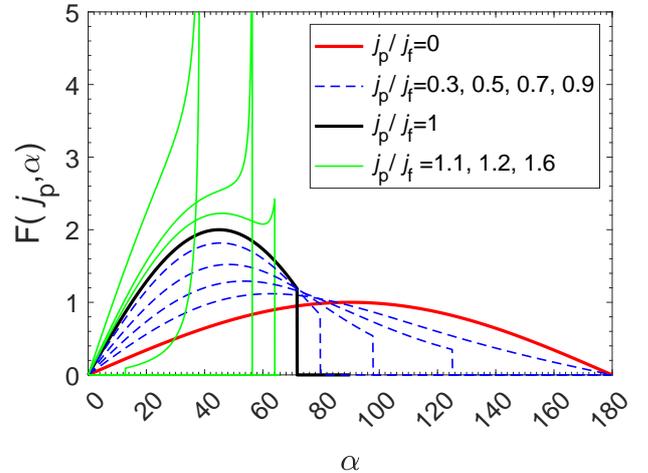} 
\caption{The distribution function according to equation (\ref{eq:Falpha}) and for $j_{\rm f}/j_{\rm L}=1.6$ (dashed-blue line in Fig. \ref{Fig:MaxAngle}).  Lines are for different values of $j_{\rm p}/j_{\rm f}$ according to the inset. In principle at one point the green lines reach infinity (where $\cos \theta=-1/\beta_p$), but the integration over $\alpha$ gives a finite value. If condition (\ref{eq:Jlimit3}) is violated $F(j_{\rm p},\alpha)=0$. For $\beta_{\rm p}=j_{\rm p}/j_{\rm f}>1$ two values of $\theta$ give the same value of $\alpha$, hence the two contributions that should be added. As far as jets are concerned values of $\alpha$ and $180^\circ - \alpha$ launch the two opposite jets along the same angle. Note that for $j_{\rm p}/j_{\rm f} \gtrsim 0.9$ no jets are launched in the equatorial plane ($\alpha=90^\circ$).   
}
\label{Fig:AlphaDistribution}
\end{figure}

In Fig. \ref{Fig:AlphaDistribution} I present the distribution function $F(j_{\rm p}, \alpha)$ from angle $\alpha=0$ to $\alpha=180^\circ$ for several values of $j_{\rm p}/j_{\rm f}$ and for one value of $j_{\rm f}/j_{\rm L}=1.6$. Note that for the launching of the jets angles $\alpha$ and $180^\circ - \alpha$ give the same jets' angle $\alpha$, as each episodes is assumed to launch two opposite jets at angles $\alpha$ and $180+\alpha$.   

Fig. \ref{Fig:AlphaDistribution} shows that as long as $j_{\rm p} \la 0.5 j_{\rm f}$ there is a large probability for jets in and near the equatorial plane ($\alpha \simeq 90^\circ$). However, when $j_{\rm p} \ga 0.9 j_{\rm f}$ the probability for jets in and near the equatorial plane is very low or zero. This further demonstrates the transition from an efficient jet feedback mechanism to an inefficient one as pre-collapse core rotation rate (as represented here by $j_{\rm p}$) increases within a relatively narrow range.

\section{Summary} 
\label{sec:Summary}

Jets that explode CCSNe can account for many of their properties, e.g., can account for the observed (e.g., \citealt{Fangetal2022} for observed stripped envelope CCSNe) common deviation from spherical symmetry of many types of CCSNe, large explosion energies (e.g., \citealt{ShishkinSoker2023}), and many more properties (for all these see the review by \citealt{Soker2022Rev}). In the present study I built (section \ref{sec:ToyModel}) and employed (sections \ref{sec:Feedback} and \ref{sec:Distribution}) a toy model to account for the $\simeq 2.5 - 5 M_\odot$ mass gap between NSs and BHs in the frame of the JJEM.

I summarised the basic properties of the toy model in Fig. \ref{Fig:Schamatic} and table \ref{TAB:Table1}. 
There are basically two sources of angular momentum of the accreted gas that forms the accretion disk or belt around the NS that launches the jets. These are the random component with a specific angular momentum amplitude of $j_{\rm f}$ and stochastically varying direction that results from pre-collapse core convection amplified by post-collapse instabilities, and the fixed specific angular momentum $j_{\rm p}$ that results from pre-collapse core rotation. 

The basic process is that when jets fully jitter, i.e., the newly born NS launches jets in all directions, as in the JJEM with zero or only slow pre-collapse core rotation, the jets expel mass from all directions during the explosion process. This terminates the accretion early on and leaves a NS remnant (e.g., \citealt{ShishkinSoker2022}). In Figs. \ref{Fig:MaxAngle} and \ref{Fig:AlphaDistribution} the cases of slow pre-collapse cores are those with $j_{\rm p}/j_{\rm f} \la 0.4$ for the minimum angular momentum that is required to launch jets of $j_{\rm L} = j_{\rm f}/1.1$, to $j_{\rm p}/j_{\rm f} \la 0.9$ for $j_{\rm L} = j_{\rm f}/2.3$. The figures show that in these cases the newly born NS can launch jets in the equatorial plane ($\alpha= 90^\circ$) and its vicinity in addition to all other directions.  

When the pre-collapse core is rapidly rotating with $j_{\rm p}/j_{\rm f} \ga 1 $ the NS does not launch jets in the equatorial plane and its vicinity. Namely, the maximum angle of the jets to the direction of $j_{\rm p}$ is $\alpha _{\rm max} < 90^\circ$. The jets in these cases might carry much more energy than in the JJEM without pre-collapse core rotation because the larger specific angular momentum of the accreted gas enables a long-lived accretion disk, rather than an intermittent one. However, the NS launches jets mainly along and near the polar directions. Accretion proceeds in the equatorial plane and adds mass to the NS to turn it into a BH. As accretion proceeds material from further out in the core that caries a larger specific angular momentum is accreted. This in turn increases the value of $j_{\rm p}/j_{\rm f}$ hence further reducing the value of $\alpha_{\rm max}$ (Fig. \ref{Fig:MaxAngle}) and as a result of that reducing the efficiency of the feedback mechanism. 

The relatively narrow range of $j_{\rm p}/j_{\rm f}$ (Fig. \ref{Fig:MaxAngle}) for the change from an efficient feedback mechanism, which leaves a NS remnant, to an inefficient one that allows the accretion to form a BH and the decrease in feedback efficiency as accretion proceeds to larger value of $j_{\rm p}$ explains the sparsely populated mass gap in the frame of the JJEM. 

\section*{Acknowledgments}

I thank Aldana Grichener, Dima Shishkin and an anonymous referee for helpful comments and Adam Soker for his help with the graphics. This research was supported by a grant from the Israel Science Foundation (769/20).

\section*{Data availability}
The data underlying this article will be shared on reasonable request to the corresponding author.  


\label{lastpage}

\end{document}